\documentclass[a4paper,reqno]{amsart}
\usepackage{setspace}
\usepackage{hyperref}
\usepackage{amsmath}
\usepackage{amsthm}
\usepackage{amssymb}
\usepackage{amsfonts}
\usepackage{mathrsfs}
\usepackage{mathtools}
\usepackage{graphicx} 
\usepackage{cite}
\usepackage{hyperref} 
\numberwithin{equation}{section}

\newcommand{\ud}{\,\mathrm{d}}
\newcommand{\mb}[1]{\boldsymbol{#1}}

\newcommand{\ms}[1]{\scriptscriptstyle{#1}}

\newcommand{\half}{\frac{1}{2}}
\newcommand{\GenVec}{\nu}
\newcommand{\Lie}{\mathscr L}
\newcommand{\eps}{\epsilon} 

\usepackage{color}

\newcounter{mnotecount}[section]

\theoremstyle{plain}
\newtheorem{theorem}{Theorem}[section]

\newtheorem{remark}[theorem]{Remark}

\title[Helicity and spin conservation in Linearized Gravity]{Helicity and spin conservation in Maxwell theory and Linearized Gravity}

\author[S. Aghapour]{Sajad Aghapour}
\address[S. Aghapour]{Albert Einstein Institute, Am M\"uhlenberg 1, D-14476 Potsdam, Germany \and Sharif University of Technology, Azadi Ave.
Tehran, Iran} 
\email{s{\textunderscore}aghapour@physics.sharif.edu}

\author[L. Andersson]{Lars Andersson}
\address[L. Andersson]{Albert Einstein Institute, Am M\"uhlenberg 1, D-14476 Potsdam, Germany }
\author[R. Bhattacharyya]{Reebhu Bhattacharyya}
\email{laan@aei.mpg.de}

\address[R. Bhattacharyya]{IIT Bombay, Powai, Mumbai, India, Postal Code: 400076}
\email{reebhu.b@iitb.ac.in}

\allowdisplaybreaks[2]

\begin{document}

\begin{abstract} 
A duality symmetric formulation of linearized gravity has been introduced by Barnett \cite{Barnett2014} and used to show the conservation of helicity. However, the relation between helicity and spin as well as the separate conservation of the spin and orbital parts of angular momentum, which is known to hold in Maxwell theory, was not considered. These conservation laws are known to follow from the conservation of the so-called helicity array, an analog of the zilch tensor, which includes helicity, spin, and spin-flux or infra-zilch.  In the present paper we prove the conservation of spin and orbital angular momentum for linearized gravity on Minkowski space, and construct the analog of the helicity array for linearized gravity. 
\end{abstract}
\date{\today \ \emph{File:} \jobname{.pdf}}
\maketitle

\section{Introduction} The Maxwell theory of electromagnetism   introduced in the mid-19th century has had a remarkably rich history, and although its modern field-theoretic formulation is simple and transparent, it is also deep and subtle. Important discoveries concerning the structure and content of Maxwell theory, including symmetries and conservation laws, as well as phenomena related to the interaction of light and matter, continue to be made. The Einstein 1915 theory of gravity, which was discovered to a large extent motivated by  the tension between Maxwell theory and Newtonian gravity, exhibits far-reaching analogies with Maxwell theory. The recent observation of gravitational waves provides a good motivation to further analyse the wave nature of Einstein theory and to explore its symmetries and conservation laws, as well as its analogies to Maxwell theory. 

Symmetries and conservation laws are fundamental features of any field theory. In the case of Maxwell equations, in addition to the Lorentz, conformal, and duality symmetries which were found by Lorentz, Heaviside, Larmor and Bateman roughly in the period 1890-1910, further non-trivial symmetries were found by Fushchich and Nikitin during the 1970's and 80's, cf. \cite{MR716301} and references therein. See Anco and Pohjanpelto \cite{MR1885280} for a classification of the local conservation laws for the Maxwell equations. Anco and The \cite{anco:the:2005:AAM:MR2220197} carried out a classification of local conservation laws for a duality symmetric formulation for Maxwell theory. The non-classical conservation laws discussed there include the zilches found by Lipkin \cite{Lipkin1964} and 
the helicity, originally found by Candlin \cite{1965NCim...37.1390C}. The variational symmetry underlying the conservation of the zilch tensor is considered in \cite{Aghapour-etal2019}.

Remarkably, a new set of conservation laws including intrinsic spin and orbital angular momentum for Maxwell theory were found in the early 1990's by Allen et al \cite{1992PhRvA..45.8185A} and van Enk and Nijenhuis \cite{1994JMOp...41..963V}. The decomposition of total angular momentum into spin and orbital angular momentum parts is well known 
but these have been viewed as not representing physical observables. The new conservation laws found in the just cited papers, which were excluded, due to locality assumptions, by the analysis of Anco et al., turn out to play an important role in experiments and their discovery has led to a burst of activity in the optical literature. These new conservation laws, which include intrinsic orbital angular momentum, spin, and spin flux or infra-zilch, were analysed in the work of Barnett et al. \cite{2010JMOp...57.1339B,Cameron2012,2012PhRvA..86a3845B} and Bliokh et al. \cite{Bliokh2014}. See also references in these papers for background. A systematic use of a duality-symmetric formulation of Maxwell theory plays a central role in this work. In particular, in the work of Barnett et al. the symmetries of the Maxwell equations giving rise to the new conservation laws via Noether's theorem were discussed.  Further, Cameron et al. have introduced an analog of Lipkin's zilch tensor, called the helicity array, cf. \cite{Cameron2012a}. The helicity array is conserved, and this property implies the conservation laws for helicity, spin and infra-zilch which were just mentioned.  

 The analogy between Maxwell theory and gravity is particularly close if we consider the weak field theory. A Maxwell-like and duality symmetric formulation for linearized gravity on Minkowski space was introduced by Barnett \cite{Barnett2014}, where the analog of helicity for linearized gravity was derived as the Noether current for the action of duality symmetry. It is worth mentioning at this point that helicity and duality symmetry for Maxwell theory and linearized gravity have previously been studied in terms of the standard formulation, and from a Hamiltonian point of view by Deser and Teitelboim \cite{1976PhRvD..13.1592D}, and Henneaux and Teitelboim \cite{PhysRevD.71.024018}.    

In this paper we shall use the duality symmetric formulation of linearized gravity introduced by Barnett in the just cited paper to derive generalizations of the helicity, spin, and infra-zilch conservation laws, and a generalization of the helicity array for linearized gravity on Minkowski space. In view of the role of spin and orbital angular momentum in the interaction of light with matter it is interesting to explore the analogous effects in gravity. 

\subsection*{Overview of this paper} In section \ref{sec:ds-em} we review the duality symmetric formulation of Maxwell theory including the conservation of helicity, cf. section \ref{subsec:helicity}, decomposition of angular momentum into its spin and orbital parts and their conservation, cf. section \ref{subsec:EM_S_cons}.  The construction of the helicity array for Maxwell theory, which contains the conservation laws for spin, and spin flux, or infra-zilch, is presented in section \ref{sec:helarr-em}.  Section \ref{GRsection} presents the duality symmetric formulation of linearized gravity on Minkowski space, and constructs the helicity array. 
The conservation of helicity for linearized gravity is presented in section \ref{sec:hel-gr}, and the decomposition of angular momentum into its spin and orbital parts is given in section \ref{sec:angmom-gr}. The helicity array for linearized gravity is presented in section \ref{sec:helarr-gr}. Section \ref{sec:conclusion} contains some concluding remarks. 
Appendix \ref{app_Symm_emtensor} gives some remarks on Belinfante-Rosenfeld symmetrization procedure and the relation between the symmetric and canonical energy-momentum tensors.

\section{Duality symmetric formulation of Maxwell theory} \label{sec:ds-em} 
\subsection{Notation and conventions} 
We shall consider fields on Minkowski space with signature $(-,+,+,+)$, using index notation with Greek indices $\alpha,\beta,\cdots$ taking values $0,\cdots,3$, and lowercase Latin indices $i,j,\cdots$ taking values $1,2,3$. Let $(x^\alpha)$ be Cartesian coordinates on Minkowski space with temporal coordinate $x^0 = t$ and spatial coordinates $(x^i)$, so that the Minkowski metric takes the form 
\begin{align} \label{eq:Minkmet}
\eta_{\alpha\beta} dx^\alpha dx^\beta = -dt^2 + \delta_{ij} dx^i dx^j , 
\end{align} 
where $\delta_{ij}$ is the Kronecker delta. For a 2-form $F_{\alpha\beta} = F_{[\alpha\beta]}$ on Minkowski space, the Hodge dual is $(*F)_{\alpha\beta} = \half \epsilon_{\alpha\beta}{}^{\gamma\delta}\,F_{\gamma\delta}$ with $\epsilon^{\alpha\beta\gamma\delta}$ the Levi-Civita symbol.
We shall sometimes make use of vector calculus notation for 3-dimensional objects, with $\mb E, \mb B$ denoting vectors with components $E^i, B^i$, respectively. We shall use natural units in which $\epsilon_0 = \mu_0 = c = 1$.

\subsection{Duality symmetric action} 
Let 
\begin{align} \label{eq:F-A}
F_{\alpha\beta} = 2\,\partial_{[\alpha} A_{\beta]} 
\end{align}
be the Faraday tensor with potential $A_\alpha$. The electric and magnetic fields $\mb E, \mb B$ for $F_{\alpha\beta}$ are
\begin{align} 
E_i = F_{i0}, \quad B_i = (*F)_{i0}
\end{align}
The 
standard Lagrangian for Maxwell theory  is, in the absence of sources,
\begin{align} 
\mathcal L_{\ms{\text{EM}}} = -\frac14\, F^{\alpha\beta}\,F_{\alpha\beta} =\frac12\,(  E^2 -   B^2) \label{standard_em_L}
\end{align}
where $E^2 = E_i E^i$, $B^2 = B_i B^i$. 
The Euler-Lagrange equation resulting from the action 
\begin{align} 
S_{\ms{\text{EM}}} = \int d^4 r \,
\mathcal L_{\ms{\text{EM}}}
\end{align} 
is   
\begin{align} 
\partial_\beta\,F^{\alpha\beta}=0
\end{align}
which together with \eqref{eq:F-A} which implies $\partial_\beta\,(*F)^{\alpha\beta}=0$ yields a form of the Maxwell equations which is manifestly invariant under the duality reflection $F_{\alpha\beta} \to (*F)_{\alpha\beta}$. 
However, the Lagrangian \eqref{standard_em_L} fails to be duality invariant. 
This motivates introducing a manifestly duality symmetric variational principle for the Maxwell equations. Let  
\begin{align}
G_{\alpha\beta} = 2\,\partial_{[\alpha} C_{\beta]}\,.
\end{align}
be an auxiliary 2-form field, which will play the role of the dual $(*F)_{\alpha\beta}$, with potential $C_\alpha$. Following \cite{Cameron2012,Bliokh2013}, we consider the duality symmetric Lagrangian  
\begin{align}
\mathcal L_{\ms{\text{EM-ds}}} = -\,\tfrac18\,(F_{\alpha\beta}F^{\alpha\beta} + G_{\alpha\beta}G^{\alpha\beta})\,. \label{EM-duplex-L}
\end{align}
with the corresponding action 
\begin{align} \label{eq:S-EM-ds}
S_{\ms{\text{EM-ds}}} = \int d^4 r \, \mathcal L_{\ms{\text{EM-ds}}}
\end{align} 
where we treat $C_\alpha$ and hence also $G_{\alpha\beta}$ as fields independent of $A_\alpha$ and $F_{\alpha\beta}$. 

The Lagrangian $\mathcal L_{\ms{\text{EM-ds}}}$ is manifestly invariant under the transformation 
\begin{align} \label{eq:F->G}
F_{\alpha\beta} \to G_{\alpha\beta}, \quad G_{\alpha\beta} \to - F_{\alpha\beta}
\end{align} 
This discrete transformation \eqref{eq:F->G} is the infinitesimal generator for the $U(1)$ action 
\begin{subequations}\label{duality_tr_F}
\begin{align}
F_{\alpha\beta}\to{}& F_{\alpha\beta}\, \cos\theta + G_{\alpha\beta}\, \sin\theta \\
G_{\alpha\beta}\to{}& G_{\alpha\beta}\, \cos\theta - F_{\alpha\beta}\, \sin\theta\,. 
\end{align}
\end{subequations}
The Euler-Lagrange equation for $\mathcal L_{\ms{\text{EM-ds}}}$ is 
\begin{align} 
\partial_\beta\,F^{\alpha\beta}=0\ ,\quad \partial_\beta\,G^{\alpha\beta}=0 \label{dual_EM_EOM}
\end{align}
Imposing the constraint 
\begin{align} 
G_{\alpha 0} = (*F)_{\alpha 0}, \quad F_{\alpha0} = - (*G)_{\alpha0}
\end{align}
on initial data at $\{t=0\}$, the solution of equations \eqref{dual_EM_EOM} has the property that the duality constraint 
\begin{align} \label{eq:duality-cond} 
G_{\alpha\beta} = (*F)_{\alpha\beta}
\end{align} 
is satisfied globally.


In the following, unless otherwise stated, we shall assume the gauge condition  
\begin{align}
A^0 = C^0 = 0\ ,\quad \nabla\cdot \mb A = \nabla\cdot \mb C = 0 .
\end{align}
This gauge condition, which we shall refer to as transverse gauge, is a combination of the temporal gauge $A^0 = C^0 = 0$, and the Coulomb gauge $\nabla \cdot \mb A = \nabla \cdot \mb C = 0$. This is consistent on Minkowski space but not on a general background. In transverse gauge, the electric and magnetic fields take the form 
\begin{subequations}\label{E&BinA&C}
\begin{align}
\mb E ={}& - \nabla \times \mb C = -\,\dot{\mb A} , \\
\mb B ={}& \nabla \times \mb A = -\,\dot{\mb C} .
\end{align}
\end{subequations}

\subsection{Helicity}
\label{subsec:helicity}
%
The Noether current for the duality symmetry \eqref{duality_tr_F} is known as the helicity current \footnote{Woltjer \cite{Woltjer489} introduced the magnetic helicity which is conserved in perfectly conducting fluids. Candlin \cite{1965NCim...37.1390C} introduced a conserved current, which he called 'screw action', that is the same as expression \eqref{eq:helicity_current} up to a trivial current, i.e. with identically vanishing divergence. Moffatt \cite{moffatt_1969} introduced the helicity of a flow in hydrodynamics and coined the term 'helicity' in classical field theory. Electromagnetic helicity was introduced first in papers by Ranada \cite{0305-4470-25-6-020,0143-0807-17-3-008}.} 
\begin{align}
J^\alpha_{\mathcal H} = \frac{\partial \mathcal L_{\ms{\text{EM-ds}}}}{\partial (\partial_\alpha A_\beta)}\,C_\beta -\frac{\partial \mathcal L_{\ms{\text{EM-ds}}}}{\partial (\partial_\alpha C_\beta)}\,A_\beta 
= \tfrac12\,(G^{\alpha\beta}\,A_\beta - F^{\alpha\beta}\,C_\beta), \label{eq:helicity_current}
\end{align}
whose components, the helicity density $\mathcal H$ and helicity flux density $\mb J_{\mathcal H}$ are, in transverse gauge
\begin{align}
J^0_{\mathcal H} \equiv \mathcal H   = \tfrac12 \left(\mb A\cdot \mb B - \mb C \cdot \mb E\right),
\quad 
\mb J_{\mathcal H} = \tfrac12\,(\mb E \times \mb A + \mb B\times \mb C )\,. \label{helicityandflux}
\end{align}
In section \ref{subsec:Em_AM}, we will show that what is found here as the helicity flux is nothing but the spin density: $\mb J_{\mathcal H} = \mb S$. Provided the equation of motion holds, the helicity current is conserved, 
\begin{subequations} 
\begin{align} 
\partial_\alpha\,J^\alpha_{\mathcal H} = 0 , \\ 
\intertext{or} 
\dot{\mathcal H} + \nabla\cdot\mb S = 0\,.\label{em_helicity_conlaw}
\end{align}
\end{subequations}

\subsection{Energy-momentum tensor} \label{sec:em-maxwell} 
%
The canonical energy-momentum tensor for $\mathcal L_{\ms{\text{EM-ds}}}$ is
\begin{align}
T_\alpha{}^{\beta} ={}& \delta_\alpha{}^{\beta}\,\mathcal L_{\ms{\text{EM-ds}}} - \frac{\partial \mathcal L_{\ms{\text{EM-ds}}}}{\partial (\partial_\beta A_\gamma)}\:\partial_\alpha A_\gamma - \frac{\partial \mathcal L_{\ms{\text{EM-ds}}}}{\partial (\partial_\beta C_\gamma)}\;\partial_\alpha C_\gamma \nonumber \\
={}& \tfrac12 \,( F^{\beta\gamma}\,\partial_\alpha A_\gamma + G^{\beta\gamma}\,\partial_\alpha C_\gamma ) \label{can_em_T}
\end{align}
If the equation of motion \eqref{dual_EM_EOM} is satisfied, the canonical energy-momentum tensor is conserved: 
\begin{align} \label{eq:T-cons-EM}
\partial_\beta\, T_\alpha{}^{\beta}=0
\end{align} 
which corresponds, via the Neother theorem, to the fact that spacetime translations are symmetries of the Lagrangian.

The components of the canonical energy-momentum tensor include the energy-momentum four-vector 
\begin{align}
    P^\alpha\equiv 
    -T_0{}^{\alpha}=(\mathcal E, \mb P), 
\end{align} 
as well as the orbital
momentum density vector $ P_{\!o\, i} = T_{i}{}^{0}$ and the canonical stress tensor $\sigma_i{}^{j} = T_i{}^{j} $. In transverse gauge, these components of the canonical energy-momentum tensor take the form 
\begin{subequations}\label{eq:emtensror_comp}
\begin{align}
    -T_{0}{}^{0} &\equiv \mathcal E  = - \tfrac12 \,(\mb E  \cdot \dot{\mb A} + \mb B \cdot \dot{\mb C}) = \tfrac12\, ( {E}^2 +   B^2) 
    \\
    -T_{0}{}^{i} &\equiv  P^i = \tfrac12\,(\mb B \times \dot{\mb A} - \mb E \times \dot{\mb C} )^i = (\mb E \times \mb B)^i 
    \\
    T_{i}{}^{0} &\equiv P_{\!o\,i} = \tfrac12\left[\,  E_j\,(\partial_i\, A^j) +   B_j\,(\partial_i\, C^j)\,\right]
    = \tfrac12 \left[\,\mb E\, .\, (\nabla) \mb A + \mb B\, .\, (\nabla)\, \mb C\,\right]_i
    \\
    T_{i}{}^{j} &\equiv \sigma_i{}^{j} = \tfrac12\,\epsilon^{jkl}\,(-  B_k\,\partial_i A_l + \,  E_k\,\partial_i C_l)\,,
\end{align}\end{subequations}
where we have used the notation 
\begin{align} 
    (\mb X\cdot(\nabla)\,\mb Y)_i = X_j \,\nabla_i\, Y^j ,\
\end{align}
cf. \cite{Berry2009}. The conservation of the energy-momentum tensor \eqref{eq:T-cons-EM} implies the conservation of energy and orbital momentum through
\begin{subequations}\begin{align}
    \partial_\alpha P^\alpha ={}& \dot{ \mathcal E} + \nabla\cdot \mb P = 0 \\
    \partial_\alpha T^\alpha_{\ i} ={}& \dot P_{\!o\,i} + \partial_j\, \sigma_i{}^j = 0
\end{align}\end{subequations}

We note that the canonical energy-momentum tensor is \emph{asymmetric}, 
\begin{align} 
    T_{\alpha\beta} \ne T_{\beta\alpha}, 
\end{align} 
The symmetric energy-momentum tensor 
\begin{align} 
    (T_s)_{\alpha\beta} = \tfrac12 (F_{\alpha\gamma} F_{\beta}{}^\gamma + G_{\alpha\gamma} G_\beta{}^\gamma ) 
\end{align} 
is related to the canonical energy-momentum tensor $T^\alpha{}_\beta$ by the Belinfante-Rosenfeld procedure, see appendix \ref{app_Symm_emtensor} for details. If the equation of motion \eqref{dual_EM_EOM} is satisfied, then the symmetric energy-momentum tensor is conserved, $\partial_\alpha (T_s)^\alpha{}_\beta = 0$, and if in addition the duality condition \eqref{eq:duality-cond} holds, we have that $(T_s)^\alpha{}_\alpha = 0$. It follows that for a conformal Killing vector $\GenVec^a$, satisfying 
\begin{align} 
    \partial_{(\alpha} \GenVec_{\beta)} - \tfrac12\, \partial_\gamma \GenVec^\gamma\, \eta_{\alpha\beta} = 0
\end{align} 
where $\eta_{\alpha\beta}$ is the Minkowski metric, 
and hence the current 
\begin{align} \label{eq:Psi-Ts}
    (\Psi_s)_\GenVec^\alpha = (T_s)^\alpha{}_\beta \,\GenVec^\beta
\end{align} 
is a conserved current, which differs by a total divergence from the Noether current corresponding to the symmetry of the action provided by the action of $\GenVec^\alpha$, cf. appendix \ref{app_Symm_emtensor}.   


The fact that the canonical energy-momentum tensor is non-symmetric is closely related to the fact that the Maxwell field has spin. A Lorentz generator $\GenVec^\alpha$ is of the form 
\begin{align} \label{eq:LorGen}
    \GenVec^\alpha = \omega^\alpha{}_\beta\, r^\beta, \quad \text{with $\omega_{\alpha\beta} = \omega_{[\alpha\beta]}$ a constant tensor.}
\end{align}   
The action on the electromagnetic field is via the Lie derivative
\begin{align} \label{eq:LieA}
    (\Lie_{\GenVec} A)^\alpha = \GenVec^\beta\, \partial_\beta A^\alpha -  A^\beta \,\partial_\beta \GenVec^\alpha
\end{align}   
provides a \emph{symmetry operator}, i.e. it takes solutions of the Maxwell equations to solutions. As we shall discuss in appendix \ref{app_Symm_emtensor}, the two terms in the right hand side of \eqref{eq:LieA} correspond to the orbital and spin parts of angular momentum.


\subsection{Angular momentum}\label{subsec:Em_AM}
The Noether current corresponding to a Lorentz generator \eqref{eq:LorGen} is of the form 
\begin{align} 
(\Psi_{\ms{\text{Noether}}})_\GenVec^\gamma = \tfrac{1}{2}\, \omega_{\alpha\beta}\, M^{\alpha\beta\gamma} 
\end{align}
where the rank-3 \emph{canonical angular momentum tensor} is anti-symmetric in the first two indices.  The anti-symmetric rank-2 tensor $M^{\alpha\beta 0}$ contains the angular momentum 3-vector $M_i=\frac12\,\epsilon_{ijk}M^{jk0}$, related to the symmetry with respect to the spatial rotations, and boost momentum 3-vector $N^i=M^{0i0}$ related to the symmetry with respect to the Lorentz boosts. It is constructive to recall that these 3-vectors are of the form $\mb M = \mb r \times \mb P$ and $\mb N = \mb P\, t - E\, \mb r$ for point particles. 

The canonical angular momentum tensor $M^{\alpha\beta\gamma}$ for the duality symmetric Lagrangian \eqref{EM-duplex-L} is given by 
\begin{align}
M^{\alpha\beta\gamma} ={}& \tilde L^{\alpha\beta\gamma} + \tilde S^{\alpha\beta\gamma}\ \label{em_can_an}
\end{align}
where its two parts come from two terms in \eqref{eq:LieA} and are
\begin{subequations}
\begin{align}
    \tilde L^{\alpha\beta\gamma} ={}&  r^\alpha T^{\beta\gamma} - r^\beta T^{\alpha\gamma} \nonumber
    \\
    ={}&  r^{[\alpha}\, (\partial^{\beta]} A^\lambda)\, F^{\gamma}{}_{\lambda}   + r^{[\alpha}\,(\partial^{\beta]} C^\lambda)\, G^{\gamma}{}_{\lambda} \label{em_L_nocons}
    \\
    \tilde S^{\alpha\beta\gamma} ={}& \frac{\partial \mathcal L_{\ms{\text{EM-ds}}}}{\partial(\partial_\gamma A^\mu)}\,(\mathcal M^{\alpha\beta})^\mu{}_\nu\,A^\nu\ + \frac{\partial \mathcal L_{\ms{\text{EM-ds}}}}{\partial(\partial_\gamma C^\mu)}(\mathcal M^{\alpha\beta})^\mu{}_\nu\,C^\nu \nonumber 
    \\
    ={}& A^{[\alpha}\,F^{\beta]\gamma}  + C^{[\alpha}\,G^{\beta]\gamma}\,. \label{em_S_nocons}
\end{align}
\end{subequations}
The total angular momentum $M^{\alpha\beta\gamma}$ is conserved,
\begin{align} 
    \partial_\gamma\, M^{\alpha\beta\gamma}=0 
\end{align} 
provided the equation of motion holds. 
We have used tilde on the two parts of the angular momentum, i.e. $\tilde L^{\alpha\beta\gamma}$ and $\tilde S^{\alpha\beta\gamma}$, to indicate that they are \textit{not} separately conserved currents, in contrast to their sum. In fact, 
\begin{align}
    \partial_\gamma\, \tilde S^{\alpha\beta\gamma}= - \partial_\gamma\, \tilde L^{\alpha\beta\gamma}= T^{\alpha\beta} - T^{\beta\alpha} \ne 0\,, \label{EM_nonconservation_eq}
\end{align}
The total angular momentum density $\mb M$ given by $M^i=\frac12\,\epsilon^i{}_{jk}\,M^{jk0}$ splits into the orbital and spin angular momentum densities $\mb L$, $\mb S$ given by \begin{align} \label{eq:LS} 
L^i={}& \tfrac12\,\epsilon^i{}_{jk} \, \tilde L^{jk0}, \quad S^i=\tfrac12\,\epsilon^i{}_{jk}\,\tilde S^{jk0},
\end{align} 
respectively. We have 
\begin{subequations} 
\begin{align}
\mb M ={}& \mb L + \mb S  
\\
\mb L ={}& \tfrac12\left[\mb E\cdot(\mb r \times \nabla)\mb A + \mb B\cdot(\mb r \times \nabla)\mb C \right] = \mb r \times \mb P_{\!o}\,,\label{orbitalAM} 
\\
\mb S ={}& \tfrac12 \left(\mb E \times \mb A + \mb B \times \mb C\right)\,. \label{spinAM}
\end{align}
\end{subequations} 
Here $\mb P_{\!o}$ is the orbital momentum density introduced in section \ref{sec:em-maxwell}. One may note that the spin density \eqref{spinAM} is exactly the same as helicity flux in \eqref{helicityandflux}.  

The two parts $\tilde L^{\alpha\beta\gamma}, \tilde S^{\alpha\beta\gamma}$ of the canonical angular momentum tensor, which contain the orbital angular momentum and spin densities respectively as their temporal components, are in fact related to the orbital and spin parts of angular momentum current, and it turns out that in spite of \eqref{EM_nonconservation_eq}, they can be modified to yield separate conservation laws which in particular encode the conservation of the orbital and spin parts of angular momentum. Moreover, as will be discussed in section \ref{sec:helarr-em}, the helicity and spin conservation laws can be encoded in the so-called helicity array $\mathcal N^{\alpha\beta\gamma}$.




\subsection{Conservation of spin and orbital parts of angular momentum}\label{subsec:EM_S_cons}
Separate conservation of orbital and spin angular momenta of free electromagnetic fields are considered by several authors, see e.g. \cite{Bialynicki-Birula2011} and \cite{2010JMOp...57.1339B}. Bliokh et al. \cite{Bliokh2014} have shown that there is a quantity $\Delta^{\alpha\beta\gamma}$, skew-symmetric in the first pair of indices, such that defining 
\begin{subequations} \label{modifiedAMcurrents}
\begin{align}
L^{\alpha\beta\gamma} ={}& \tilde L^{\alpha\beta\gamma} + \Delta^{\alpha\beta\gamma}, \\ S^{\alpha\beta\gamma} ={}& \tilde S^{\alpha\beta\gamma} - \Delta^{\alpha\beta\gamma},
\end{align} 
\end{subequations} 
we have 
\begin{align} 
\partial_\gamma L^{\alpha\beta\gamma}= \partial_\gamma S^{\alpha\beta\gamma} = 0 
\end{align}
Further, as we shall see, $\Delta^{\alpha\beta\gamma}$ can be chosed such that $\Delta^{\alpha\beta0}=0$, and hence 
\begin{align} \label{eq:LS-mod}  
L^i = \half \eps^i{}_{jk} L^{jk0} = \half \eps^i{}_{jk} \tilde L^{jk0}, \quad 
S^i = \half \eps^i{}_{jk} S^{jk0} = \half \eps^i{}_{jk} \tilde S^{jk0} 
\end{align} 
so that the modified orbital angular momentum, and spin densities agree with those given in \eqref{eq:LS}. 


From \eqref{em_S_nocons}, $\tilde S^{ijk}$ is anti-symmetric in two first indices. Its dual satisfies 
\begin{align}
 \tfrac12\,\epsilon^{ijk}\,\partial_\gamma \tilde S_{jk}{}^{\gamma}= \dot S^i + \partial_j\, \tilde \Sigma^{ij} = \epsilon^i{}_{jk}\,T^{jk}
\end{align}
in which, the \textit{false} spin flux $\tilde \Sigma_{ij}$ obtained from nonconserved spin current $\tilde S^{\alpha\beta\gamma}$, is
\begin{align}
\tilde \Sigma_{ij} ={}& \tfrac12\,\epsilon_{i}{}^{kl}\,\tilde S_{klj} \nonumber \\ 
={}& \tfrac12\left[\,\delta_{ij}\, (\mb A\cdot\mb B - \mb C\cdot\mb E) + E_i\,C_j - B_i\,A_j\, \right] 
\nonumber\\
={}& \delta_{ij}\,\mathcal H + \tfrac12\,( E_i\,C_j - B_i\,A_j ), \label{em_Sflux_false}
\end{align}
where $\mathcal H$ is the helicity of electromagnetic field \eqref{helicityandflux}. 

In order to have a proper continuity equation for the spin density $\mb S$, the right hand side of equation \eqref{em_Sflux_false} should be absorbed in the flux term. This can be done by finding a modifying term $\Delta_{ij}$ satisfying 
\begin{align}
\partial_j\,\Delta_{i}{}^j ={}& \epsilon_{ijk}\,T^{jk} \nonumber \\
={}& \tfrac12\,\epsilon_{i}{}^{jk}\,\epsilon_{k}{}^{lm}\left(  B_m\,\partial_j\,A_l - E_m\,\partial_j\,C_l \right) \nonumber \\ 
={}& \tfrac12\,\partial_j\left(B^j\,A_i - E^j\,C_i\right).
\end{align}
This suggests that $\Delta_{ij} = \frac12\,(B_j\,A_i - E_j\,C_i)$ and the modified \textit{true} spin flux\footnote{Cameron et al. in \cite{Cameron2012a} use the term infra-zilch for the spin flux and denote it by $n_{ij}$.} is
\begin{align}
\Sigma_{ij} = \tilde\Sigma_{ij} - \Delta_{ij} = \delta_{ij}\,\mathcal H + E_{(i}\,C_{j)} - B_{(i}\,A_{j)} \,. \label{em_Sflux}
\end{align}
One may note that the above modification leads to the symmetrization of spin flux. The conservation law for spin takes the form 
\begin{align}
\dot S_i + \partial_j\, \Sigma_i{}^{j} = 0\,. \label{em_S_continuity}
\end{align}
In order to verify this, we calculate the first term in \eqref{em_S_continuity}, 
\begin{align}
    \dot {\mb S} ={}& \tfrac12\,(\dot{\mb E}\times\mb A + \dot{\mb B}\times\mb C) \nonumber \\ ={}& \tfrac12\left[(\nabla\times\mb B)\times \mb A - (\nabla\times \mb E)\times \mb C\right]\, \label{dot_S}
\end{align}
where for the second equality we have use the Maxwell equations. Next, we calculate the divergence of spin flux (second term in \eqref{em_S_continuity}) and show that it is indeed negative of \eqref{dot_S},
\begin{align}
    \partial_j\,\Sigma_i{}^j ={}& \partial_i \mathcal H + \tfrac12 \left[ (\mb C\cdot\nabla)\mb E + (\mb E\cdot \nabla)\mb C - (\mb A\cdot\nabla)\mb B - (\mb B\cdot \nabla)\mb A\right]_i\nonumber
    \\
    ={}& \tfrac12 \left[\mb A \times (\nabla\times\mb B) + \mb B \times (\nabla\times\mb A) - \mb C \times (\nabla\times\mb E) - \mb E \times (\nabla\times\mb C)\right]_i\label{div_of_spin_flux}
\end{align}
Here we have used the identity
\begin{align}
    \nabla \mathcal (\mb A \cdot \mb B) = \mb A \times (\nabla\times\mb B) + \mb B \times (\nabla\times\mb A) + (\mb A\cdot\nabla)\mb B + (\mb B\cdot \nabla)\mb A\,.
\end{align}
The second and forth terms in \eqref{div_of_spin_flux} are zero and the remaining terms cancel that in \eqref{dot_S}. This proves that the conservation law for spin \eqref{em_S_continuity} holds.

The modifying term $\Delta_{ij}=\frac12\,\epsilon_{i}{}^{kl}\,\Delta_{klj}$ of spin flux is a component of the 3-index modification term $\Delta^{\alpha\beta\gamma}$ in \eqref{modifiedAMcurrents}. All components of the 3-index modification tensor can be obtained by comparing equations \eqref{EM_nonconservation_eq} and \eqref{modifiedAMcurrents}, which suggests the relation $\partial_\gamma \Delta^{\alpha\beta\gamma} = T^{\alpha\beta} - T^{\beta\alpha}$, and noting that we wish to carry out this modification in such a manner that the orbital and spin angular momentum densities $\mb L$ and $\mb S$ in \eqref{orbitalAM} and \eqref{spinAM} are not affected. Based on these considerations, the components of $\Delta^{\alpha\beta\gamma}$ in transverse gauge can be defined as 
\begin{subequations}
\begin{align}
\Delta^{\alpha\beta 0} ={}& \Delta^{00\gamma} = 0, \\
\Delta^{i0k} ={}& - \Delta^{0ik} = \tfrac12(A^i\, E^k + C^i\, B^k), \\
\Delta^{ijk} ={}& \tfrac12\,\epsilon^{ij}{}_l\,(A^l\,B^k - C^l\, E^k) \label{EM_Delta}
\end{align}
\end{subequations}
With this choice of $\Delta^{\alpha\beta\gamma}$, \eqref{modifiedAMcurrents} yields the conserved orbital and spin angular momentum currents $L^{\alpha\beta\gamma}$ and $S^{\alpha\beta\gamma}$. The spin conservation law, 
\begin{align} 
\partial_\gamma S^{\alpha\beta\gamma}=0, 
\end{align} 
leads to the continuity relation \eqref{em_S_continuity} for the spin density vector $S_i$.
Similarly, the orbital angular momentum conservation law, $\partial_\gamma L^{\alpha\beta\gamma}=0$, leads to the continuity relation 
\begin{align}
\dot L_i + \partial_j \Lambda_{i}{}^j ={}& 0 , 
\end{align}
for the orbital angular momentum density vector, which in view of \eqref{eq:LS-mod} is  
as in \eqref{orbitalAM}, and its flux density $\Lambda_{ij}=\frac12\, \epsilon_{i}{}^{kl}\,L_{klj}$ given by 
\begin{align}  
\Lambda_{ij} ={}& \tfrac12 \left\{\epsilon_{ikl}\,\epsilon_{jmn}\,r_k\left[  B_n\,(\partial_l\,A_m) - E_n\,(\partial_l\,C_m)\right] +A_i\,  B_j - C_i\,  E_j\right\}
\end{align}


\subsection{Helicity array} \label{sec:helarr-em}
We found in subsection \ref{subsec:helicity} that the continuity equation for helicity $\mathcal H$ contains the spin part of angular momentum $\mb S$ as the flux of helicity, cf. equation \eqref{helicityandflux}. Further, the spin part of angular momentum itself is in fact conserved and obeys a continuity equation with flux $\Sigma_{ij}$, cf.  \eqref{em_Sflux}. In fact, the spin flux is also conserved. We can find the continuity equation directly by calculating the time derivative of spin flux in transverse gauge,
\begin{align}
\dot{\Sigma}_{ij} ={}& \delta_{ij}\,\dot{\mathcal{H}} - \dot B_{(i}\,A_{j)} - B_{(i}\,\dot A_{j)} + \dot E_{(i}\,C_{j)} + E_{(i}\,\dot C_{j)} 
\nonumber\\
={}& \delta_{ij}\,(-\partial_k\,S^k) + (\nabla\times E)_{(i}\,A_{j)} + (\nabla\times B)_{(i}\,C_{j)}
\nonumber\\
={}& \delta_{ij}\,(-\partial_k\,S^k) + \nabla^2 C_{(i}\,A_{j)} - \nabla^2 A_{(i}\,C_{j)}
\nonumber\\
={}&  \partial_k \left(-\delta_{ij}\,S^k + A_{(i}\,\partial^k C_{j)} - C_{(i}\,\partial^k A_{j)} \right).
\end{align}
In second line, we have used the Maxwell equations as well as \eqref{E&BinA&C}. In third line, we have used the identity $\nabla\times(\nabla \times \mb A) = \nabla(\nabla\cdot\mb A) - \nabla^2 \mb A$ and the same for $\mb C$ and in the last line, an integration by parts and the transverse conditions for potentials. 
Thus the flux of spin flux (or flux of infra-zilch in the terminology of \cite{Cameron2012a}) is
\begin{align}
N^{ij}{}_k = \delta^{ij}\,S_k - A^{(i}\,\partial_{k} C^{j)} + C^{(i}\,\partial_{k} A^{j)}
\end{align}
and the continuity equation for $\Sigma_{ij}$ is 
\begin{align}
\dot{\Sigma}_{ij} + \partial_k \, N_{ij}{}^k = 0\,. \label{em_infrazilich_convlaw}
\end{align}
Motivated by the similarity of the conservation laws \eqref{em_helicity_conlaw}, \eqref{em_S_continuity} and \eqref{em_infrazilich_convlaw}, Cameron et al. in \cite{Cameron2012a} arranged these quantities in the 3-index \emph{helicity array} $\mathcal N^{\alpha\beta\gamma}$, with symmetry $\mathcal N^{\alpha\beta\gamma} = \mathcal N^{(\alpha\beta)\gamma}$. The helicity array has 27 independent components given by  \begin{subequations}
\begin{align}
\mathcal N^{000} &\equiv \mathcal H = \tfrac12\,(\mb A\cdot\mb B - \mb C\cdot\mb E)\,,
\\
\mathcal N^{0i0} &= \mathcal N^{00i} \equiv S^i = \tfrac12\,(\mb E \times \mb A + \mb B \times \mb C)^i\,,
\\
\mathcal N^{ij0} &= \mathcal N^{0ij} \equiv \Sigma^{ij} = \delta^{ij}\,\mathcal H + E^{(i}\,C^{j)} - B^{(i}\,A^{j)}\,,
\\
\mathcal N^{ij}{}_k &\equiv N^{ij}{}_k = \delta^{ij}\,S_k - A^{(i}\,\partial_{k} C^{j)} + C^{(i}\,\partial_{k} A^{j)}\,,
\end{align}
\end{subequations}
Despite its suggestive structure, $\mathcal N^{\alpha\beta\gamma}$ it is not a tensor. The significance of this helicity array is in the fact that all conservation laws \eqref{em_helicity_conlaw}, \eqref{em_S_continuity} and \eqref{em_infrazilich_convlaw} for helicity, spin and spin flux, or infra-zilch, are contained in the continuity relation
\begin{align}
\partial_\gamma\,\mathcal N^{\alpha\beta\gamma}=0\,.
\end{align}
For ease of reference, we restate these here, 
\begin{subequations}\label{eq:allcons}
\begin{align}
    &\dot{\mathcal H} + \partial_i \, S^i = 0
    \\
    &\dot{S}_{i} + \partial_j \, \Sigma_i^{\ j} = 0
    \\
    &\dot{\Sigma}_{ij} + \partial_k \, N_{ij}^{\;\ k} = 0
\end{align}
\end{subequations}
As pointed out in \cite{Cameron2012a} the transformation 
\begin{align}
    \mb A \to \nabla\times \mb A\ ,\quad \mb C\to \nabla\times \mb C\,,
\end{align}
takes the helicity array to Lipkin's Zilch tensor \cite{Lipkin1964}
\begin{subequations}\label{zilch_comps}
\begin{align}
    Z^{000} ={}& \tfrac12\,(\mb B\cdot (\nabla\times \mb B) + \mb E \cdot (\nabla \times \mb E)\,,
    \\
    Z^{0i0} ={}& Z^{00i}= \tfrac12\left[\,(\nabla\times \mb E)\times \mb B - (\nabla\times \mb B)\times \mb E\,\right]^i\,
    \\
    Z^{ij0} ={}& Z^{0ij}= \delta^{ij}\,Z^{000} -(\nabla\times\mb E)^{(i}\, E^{j)} - (\nabla\times\mb B)^{(i}\, B^{j)}\,
    \\
    Z^{ij}{}_k ={}& \delta_{ij}\,Z^{00}{}_k + B^{(i}\,\partial_{k}E^{j)} - E^{(i}\,\partial_{k}B^{j)}\,,
\end{align}\end{subequations}
This transformation may be iterated to yield the higher order zilches. The zilch tensor can be written in a covariant form
\begin{align}
    Z^{abc} = F^{cd}\,{*F}_d{}^{(a,b)} - \,*F^{cd}\,F_d{}^{(a,b)}
\end{align}
which is equivalent (up to a trivial current) to the set components in \eqref{zilch_comps}. This covariant form is manifestly symmetric between two first indices but for solutions of Maxwell equations, it is totally symmetric. The variational symmetry underlying the conservation of the zilch tensor is considered in \cite{Aghapour-etal2019} via the inverse Noether procedure.

\section{Duality symmetric formulation of linearized gravity}\label{GRsection}

\subsection{Background and notation} 
Barnett \cite{Barnett2014} exploited the analogy of gravity with Maxwell theory to introduce a duality-symmetric formulation of linearized gravity on Minkowski space, and used this to derive the helicity of the gravitational field. %
Truncating the Einstein-Hilbert Lagrangian
\begin{align}
    \mathcal L_{\ms{\text{EH}}} = \frac{\sqrt{-g}}{16\pi G}\, \mathcal R\,.
\end{align}
yields a Lagrangian for linearized gravity which, after adding a total derivative and setting $32\pi G =1 $, takes the form  
\begin{align}
    \mathcal L_{\ms{\text{LG}}} = \tfrac12\, (\partial_\beta h^\alpha{}_\alpha \,\partial^\beta h^\gamma{}_\gamma -2\,\partial_\beta h^\alpha{}_\alpha \,\partial^\gamma h^\beta{}_\gamma - \partial_\gamma h_{\alpha\beta}\, \partial^\gamma h^{\alpha\beta} + 2\, \partial_\gamma h_{\alpha\beta}\,\partial^\beta h^{\alpha\gamma} )
\end{align}
where $h_{\alpha\beta} = h_{(\alpha\beta)}$ is the linearized metric. In the gauge 
\begin{align}
    h_{0\alpha}=0 \ ,\quad  h_{ij}{}^{,j}= 0\ ,\quad h^i{}_{i} = 0\,.
\end{align}
which may be consistently imposed on Minkowski space, the Lagrangian $\mathcal L_{\ms{\text{LG}}}$ takes the form
\begin{align}
    \mathcal L_{\ms{\text{LG}}} = \tfrac12 \,(\dot h_{ij}\,\dot h^{ij} - h_{ij,k}\,h^{ij,k} + 2\, h_{ij,k}\,h^{ik,j} )\,.
\end{align}
Adding a total derivative term gives the dynamically equivalent Lagrangian
\begin{align}
    \mathcal L_{\ms{\text{LG}}}' ={}& \mathcal L_{\ms{\text{LG}}} - \tfrac12\,(h_{jk}\,h^{ik,j})_{,i} \nonumber \\ 
    ={}& \tfrac12 \,(\dot h_{ij}\,\dot h^{ij} - h_{ij,k}\,h^{ij,k} + \, h_{ij,k}\,h^{ik,j} )
\label{Lprime}
\end{align}


We shall make use of analogues of vector calculus operations for symmetric 2-tensors here, and now introduce the notation which will be used. Let $c_{ij}$, $d_{ij}$ be symmetric 2-tensors. We use bold faced letters $\mb c$ for a symmetric, traceless 2-tensor like $c_{ij}$. Further, we shall use several binary operations. These are 
\begin{subequations}\begin{align}
    &\text{the scalar dot product} 
    &&\mb c \cdot \mb d = c_{ij}\,d^{ij}\,,
    \\
    &\text{the cross product} 
    &&(\mb c \times \mb d)_i = \epsilon_{i}{}^{jk}\,c_{jl}\,d_{k}{}^{l}\,,
    \\
    &\text{2-tensor dot product} 
    &&(\mb c\, : \,\mb d)_{ij} = c_{k(i}\,d_{j)}{}^k\,,
    \\
    &\text{and the wedge product} 
    &&(\mb c \wedge \mb d)_{ij} = \epsilon_i{}^{kl}\,\epsilon_j{}^{mn}\,c_{km}\,d_{ln}\,.
\end{align}\end{subequations}
We also define the divergence and curl of a symmetric 2-tensor $e_{ij}$ as
\begin{align}
    (\nabla\cdot\mb e)_i = e_{ij}{}^{,j}\ , \quad (\nabla \times \mb e)_{ij} = \epsilon_{(i}^{\;\ kl}\,e_{j)l,k} &
\end{align}

The symmetric, traceless 2-tensor fields 
\begin{align}
    e_{ij}= -\dot h_{ij}\ ,\quad b_{ij} = \epsilon_{i}^{\ lm}\,h_{jm,l}
\end{align}
will play the role of analogues of the electric and magnetic fields in Maxwell theory. Using the notation we have just introduced, we have 
\begin{align} 
    \mathcal L_{\ms{\text{LG}}}' 
    ={}& \tfrac12 \left[\dot{\mb h}\cdot \dot{\mb h} - (\nabla\times \mb h)\cdot(\nabla\times \mb h)\right] \nonumber \\ 
    ={}& \tfrac12\, (\mb e \cdot \mb e - \mb b \cdot \mb b)\,, 
\end{align} 
which is closely analogous to the standard Lagrangian \eqref{standard_em_L} for Maxwell theory. The Euler-Lagrange equations which follow from this Lagrangian, when written in our current notation, take the form
\begin{align}
    &\nabla\cdot\mb e=0\ , &&\nabla\cdot\mb b =0\ , &&\nabla \times \mb e = -\dot{\mb b}\ , &&\nabla \times \mb b = \dot{\mb e}\,.
\end{align}
which is close to the free Maxwell equations. 

In order to construct a duality symmetric Lagrangian for linearized gravity, we introduce a second (auxiliary) gravitational potential $k_{ij}$, which is analogue of 4-potential $C^\alpha$ in duality symmetric Maxwell theory and impose the transverse-traceless gauge conditions on it. 

\subsection{Duality symmetric Lagrangian} 
Let $h_{ij}$, $k_{ij}$ be a pair of symmetric, traceless $2$-tensors. Following the procedure that was used in the Maxwell case, we introduce a duality symmetric Lagrangian for linearized gravity
\begin{align}
    \mathcal L_{\ms{\text{LG-ds}}} = \tfrac14\left[\dot{\mb h}\cdot \dot{\mb h} - (\nabla\times \mb h)\cdot(\nabla\times \mb h)
    + \dot{\mb k}\cdot \dot{\mb k} - (\nabla\times \mb k)\cdot(\nabla\times \mb k)\right]  . \label{ds_L_LG}
\end{align}
This is manifestly invariant the duality reflection 
\begin{align} 
h_{ij} \to{}& k_{ij}, \quad k_{ij} \to - h_{ij}
\end{align}
which generates the continuous $U(1)$ duality rotation 
\begin{align}
    h_{ij}\to h_{ij}\,\cos\theta + k_{ij}\,\sin\theta\ , \quad k_{ij}\to k_{ij}\,\cos\theta - h_{ij}\,\sin\theta \,. \label{eq:gr-dualtrans}
\end{align}
Note that, similar to the duality symmetric Lagrangian of Maxwell theory, this Lagrangian is zero under the duality constraint, 
\begin{subequations}
\begin{align} 
\dot{\mb h} ={}& \nabla \times \mb k \\ 
\dot{\mb k} ={}& - \nabla \times \mb h
\end{align} 
\end{subequations}
It follows from the Euler-Lagrange equations for $\mathcal L_{\ms{\text{LG-ds}}}$ that the duality constraint holds globally if it holds at $t=0$. In the following, unless otherwise stated, we shall impose the duality constraint and the transverse traceless gauge condition, 
\begin{subequations} 
\begin{align} 
h_{ij}{}^{,j} ={} 0, \quad h^i{}_i = 0 \\
k_{ij}{}^{,j} ={} 0, \quad k^i{}_i = 0 .
\end{align} 
\end{subequations}
The Maxwellian gravitational fields $e_{ij}$ and $b_{ij}$ in terms of these potentials are \begin{subequations}\label{eandb}
\begin{align}
    \mb e ={}& - \dot{\mb h} = - \nabla\times \mb k
    \\
    \mb b ={}& - \dot{\mb k} = \nabla\times\mb h\,.
\end{align}\end{subequations}

\subsection{Helicity}
\label{sec:hel-gr}
Invariance of the Lagrangian density \eqref{ds_L_LG} under the duality transformation \eqref{eq:gr-dualtrans}
leads to the conservation of the current
\begin{align}
    J^\alpha = \frac{\partial \mathcal L_{\ms{\text{LG-ds}}} }{\partial (h_{ij,\alpha})}\,k_{ij} - \frac{\partial \mathcal L_{\ms{\text{LG-ds}}} }{\partial (k_{ij,\alpha})}\,h_{ij} \ ,\quad \partial_\alpha\,J^\alpha = 0\,,
\end{align}
whose components in transverse-traceless gauge are
\begin{subequations}\begin{align}
    J^0 ={}& \tfrac12\, (\dot h_{ij} k^{ij} - \dot k_{ij} h^{ij}) = \tfrac12\, (\mb h\cdot\mb b - \mb k\cdot\mb e)
    \\
    J^i ={}& \tfrac12\,\epsilon^{ijk}(e_{lj}\,h^l{}_{k} + b_{lj}\,k^l{}_{k})= \tfrac12\,(\mb e\times \mb h +\mb b \times \mb k)_k \,.
\end{align}\end{subequations}
This current is analogous to the electromagnetic helicity current \eqref{helicityandflux}, but it will be shown later that the flux $J^k$ is one half of the spin density vector $S^k$, which is the consequence of describing the gravitational field by a symmetric 2-tensor. Thus we define the helicity $\mathcal H$ and spin $\mb S$ of the gravitational field to be\footnote{This definition of gravitational helicity differs from which is defined in \cite{Barnett2014} by a factor of 2.}
\begin{align}
    \mathcal H &\equiv 2\, J^0 = \mb h\cdot\mb b - \mb k\cdot\mb e \\
    \mb S &\equiv 2\, \mb J = \mb e\times \mb h +\mb b \times \mb k
\end{align}
The helicity conservation law is
\begin{align}
    \dot{\mathcal H} + \nabla\cdot\mb S = 0 \,.
\end{align}

\subsection{Energy-momentum tensor}
The canonical energy-momentum tensor for the duality symmetric Lagrangian \eqref{ds_L_LG} is
\begin{align}
    T_{\alpha}{}^{\beta} = \delta_{\alpha}{}^{\beta}\,\mathcal L_{\ms{\text{LG-ds}}} - \frac{\partial \mathcal L_{\ms{\text{LG-ds}}}}{\partial (h_{ij,\beta})}\;h_{ij,\alpha} - \frac{\partial \mathcal L_{\ms{\text{LG-ds}}}}{\partial ( k_{ij,\beta})}\;k_{ij,\alpha} \ ,\quad \partial_\beta\, T_{\alpha}{}^{\beta}=0\,,
\end{align}
whose components in transverse-traceless gauge are
\begin{subequations}\begin{align}
    -T_{0}{}^{0} &\equiv \mathcal E = \tfrac12\,(\dot h_{ij}\dot h^{ij} + \dot k_{ij} \dot k^{ij}) = \tfrac12\,(\mb e\cdot\mb e + \mb b\cdot\mb b) 
    \\
    -T_0{}^{i} &\equiv P^i = - \tfrac12 \,(\epsilon^{ijk}\,b_{nj}\,\dot h^n_{\ k} - \epsilon^{ijk}\,e_{nj}\dot k^n_{\ k}) = ( \mb e \times \mb b)_i
    \\
    T_i{}^{0} &\equiv P_{\!o}^i = - \tfrac12 \,(\dot h^{jk}\,h_{jk,i} + \dot k^{jk}\,k_{jk,i}) = \tfrac12 \left[ \mb e\cdot (\nabla)\mb h + \mb b\cdot (\nabla)\mb k \right]_i
    \\
    T_i{}^{j} &\equiv \sigma_i{}^j= \tfrac12\,(-\epsilon^{jkl}\,b^n{}_{k}\,h_{nl,i} + \epsilon^{jkl}\,e^n{}_{k}\,k_{nl,i} )
\end{align}\end{subequations}
where we have used the notation $\left[\mb e \cdot (\nabla)\mb h\right]_i = e^{jk}\,h_{jk,i}$ similar to the notation in previous section.

\subsection{Angular momentum}\label{sec:angmom-gr}
Similar to Maxwell theory (section \ref{subsec:Em_AM}), the canonical angular momentum current in linearized gravity can be derived as
\begin{align}
    M^{\alpha\beta\gamma} ={}& \tilde L^{\alpha\beta\gamma} + \tilde S^{\alpha\beta\gamma}\ , \quad \partial_\alpha\, M^{\alpha\beta\gamma}=0 \label{lg_can_an}
\end{align}
where nonconserved orbital and spin angular momenta, $\tilde L^{\alpha\beta\gamma}$ and $\tilde S^{\alpha\beta\gamma}$, are
\begin{subequations}\begin{align}
    \tilde L^{\alpha\beta\gamma} ={}&  r^\alpha T^{\beta\gamma} - r^\beta T^{\alpha\gamma}
    \\
    \tilde L^{ij0} ={}& e_{kl}\,h^{kl,[j}\,r^{i]} + b_{kl}\,k^{kl,[j}\,r^{i]}
    \\
    \tilde L^{ijk} ={}&   \epsilon_{lmn}\,b_p^{\;m}\,h^{pn,[i}\,r^{j]} + \epsilon_{lmn}\,e_p^{\;m}\,k^{pn,[i}\,r^{j]}
\end{align}\end{subequations}
and
\begin{align}
    \tilde S_{ij}{}^{\gamma} ={}& \frac{\partial \mathcal L_{\ms{\text{LG-ds}}}}{\partial(\partial_\gamma h_{mn})}\left[(\mathcal M_{ij})_{ml}\,h_{ln} + (\mathcal M_{ij})_{nl}\,h_{ml}\right]+ \frac{\partial \mathcal L_{\ms{\text{LG-ds}}}}{\partial(\partial_\gamma k_{mn})}\left[(\mathcal M_{ij})_{ml}\,k_{ln} + (\mathcal M_{ij})_{nl}\,k_{ml}\right] 
    \nonumber\\
    \tilde S_{ij}{}^{0}={}& \tfrac12 \,(e_{n[i}\,h_{j]n} +b_{n[i}\,k_{j]n} )
    \nonumber\\
    \tilde S_{ij}{}^{k}={}& \tfrac14 \,( \epsilon_{kln}\,h_{l[i}\,b_{j]n}  - \epsilon_{kl[i}\,h_{j]n}\,b_{nl} - \epsilon_{kln}\,k_{l[i}\,e_{j]n}  + \epsilon_{kl[i}\,k_{j]n}\,e_{nl})
    \label{LG_spin_current}
\end{align}
and the spin density 3-vector is obtained as
\begin{align}
    S_i = \tfrac12\,\epsilon_{ijk}\,\tilde S^{jk0} = \epsilon_{ijk}\,(e^{mj}\,h_m{}^k +b^{mj}\,k_m{}^k ) = (\mb e\times \mb h + \mb b \times \mb k)_i \,, \label{eq:lg_s_density}
\end{align}
Similarly to the Maxwell case, the conservation of spin and orbital angular momenta of the gravitational field cannot be found directly from $\tilde S^{\alpha\beta\gamma}$, $\tilde L^{\alpha\beta\gamma}$ which are not separately conserved. We have 
\begin{align}
    \partial_\gamma\, \tilde S^{\alpha\beta\gamma}= - \partial_\gamma\, \tilde L^{\alpha\beta\gamma}= T^{\alpha\beta} - T^{\beta\alpha} \ne 0\,, \label{LG_nonconservation_eq}
\end{align}
To find the proper conservation laws, we need to modify the spin and orbital angular momentum fluxes in a way that total angular momentum conservation remains unchanged.

The false spin flux obtained from nonconserved spin current $\tilde S^{\alpha\beta\gamma}$ in \eqref{LG_spin_current} is
\begin{align}
\tilde \Sigma_{ij} = \tfrac12\, \epsilon_{ikl}\,\tilde S^{kl}{}_j = \delta_{ij}\,\mathcal H - b_{ki}\,h^k{}_j - \tfrac12\,b_{kj}\,h^k{}_i + e_{ki}\,k^k{}_j + \tfrac12\,e_{kj}\,k^k{}_i
\end{align}
Whit similar analysis to the one we made for Maxwell case, the modifying term $\Delta^{\alpha\beta\gamma}$ can be find to have the components, in transverse-traceless gauge, as
\begin{subequations}\begin{align}
&\Delta^{\alpha\beta 0} = \Delta^{00\gamma} = 0\,
\\
&\Delta^{i0j} = - \Delta^{0ij} = \tfrac12\,( h_{ki}\, e^k{}_{j} + k_{ki}\, b^k{}_{j})\,,
\\
&\Delta^{ijk} = \tfrac12\,\epsilon^{ijl}\,(h_{ml}\,b^m{}_{k} - k_{ml}\,e^m{}_{k}) \label{LG_Delta}
\end{align}\end{subequations}
The first of these three equations is the condition for not altering the spin and orbital angular momentum densities. The second one modifies the boost angular momentum flux and the third one modifies the spin and orbital angular momentum fluxes. The latter modification results in (symmetric) spin flux
\begin{align}
\Sigma_{ij} = \tfrac12\, \epsilon_{ikl}\,(\tilde S^{kl}{}_j - \Delta^{kl}{}_j) = \delta_{ij}\,\mathcal H + 2\, e_{k(i}\,k_{j)}{}^k - 2\, b_{k(i}\,h_{j)}{}^k\,,
\end{align}
which together with the spin density \eqref{eq:lg_s_density} satisfies the continuity relation
\begin{align}
\dot S_i + \partial_j\,\Sigma_{i}{}^j = 0\,.
\end{align}

\subsection{Helicity array for linearized gravity}\label{sec:helarr-gr}
Similar to the Maxwell case, the spin flux of linearized gravitational field is conserved:
\begin{align}
\dot \Sigma_{ij} + \partial_k\,N_{ij}{}^k =0
\end{align}
where the flux $N_{ijk}$ can be obtained easily by computing the time derivative of $\Sigma_{ij}$ and observe that it is in fact a total derivative. This results in
\begin{align}
N_{ijk} = \delta_{ij}\,S_k -2\, h_{n(i}\,k_{j)n,k} + 2\, k_{n(i}\,h_{j)n,k}
\end{align}

The helicity array, then, can be constructed as
\begin{subequations}
\begin{align}
\mathcal N^{000} &\equiv \mathcal H = \mb h\cdot\mb b - \mb k\cdot\mb e\,,
\\
\mathcal N^{0i0} &= \mathcal N^{00i} \equiv S^i = (\mb e \times \mb h + \mb b \times \mb k)^i\,,
\\
\mathcal N^{ij0} &= \mathcal N^{0ij} \equiv \Sigma^{ij} = \delta^{ij}\,\mathcal H + 2\, e_{k}{}^{(i}\,k^{j)k} - 2\, b_{k}{}^{(i}\,h^{j)k}\,,
\\
\mathcal N^{ijk} &\equiv N^{ijk} = \delta^{ij}\,S^k -2\, h_{l}{}^{(i}\,k^{j)l,k} + 2\, k_{l}{}^{(i}\,h^{j)l,k}\,,
\end{align}\end{subequations}
with the symmetry $\mathcal N^{\alpha\beta\gamma}=\mathcal N^{(\alpha\beta)\gamma}$. The continuity relation 
\begin{align}
    \partial_\gamma \, \mathcal N^{\alpha\beta\gamma} = 0
\end{align}
contains the linked conservation laws 
\begin{subequations}\begin{align}
    &\dot{\mathcal H} + \partial_i S^i = 0,
\\
    &\dot S_i + \partial_j \Sigma_i^{\ j} =0,
\\
    &\dot \Sigma_{ij} + \partial_k N_{ij}^{\;\ k} =0
\end{align} \end{subequations}

\begin{remark} 
In the case of Maxwell theory, the transformation
\begin{align} 
\mb A \to{}& \nabla\times A \\ 
\mb C \to{}& \nabla \times C
\end{align} 
takes the Helicity array to the Zilch tensor. It is reasonable to conjecture that this holds also in the case of linearized gravity. 
\end{remark} 
\section{Conclusion} \label{sec:conclusion}
We have presented the duality symmetric formulation of linearized gravity on Minkowski space, and derived the generalization from Maxwell theory of the conservation laws for helicity, spin, and infra-zilch to the gravitational case. 

The fact that the spin and orbital parts of angular momentum are separately conserved and therefore physical observables has had a tremendous impact in optics and in our understanding of the interaction of light and matter. 
Here we have shown that the spin and orbital parts of angular momentum are separately conserved also in linearized gravity on Minkowski space. It is now interesting to analyze the consequences of this fact for the interaction of gravity with matter as well as with other fields. We remark that Bialynicki-Birula and Bialynicki-Birula \cite{Bialynicki-Birula2016} have constructed beams of gravitational waves carrying orbital angular momentum. Recently \cite{2018arXiv181002219B}, the interaction of such beams with matter has been investigated. 

The analysis presented here relies on the transverse-traceless gauge. It has been shown by Bialynicki-Birula and Bialynicki-Birula \cite{Bialynicki-Birula2011} that a gauge-invariant, but non-local, expression for spin and orbital angular momentum can be given by using the Biot-Savart law. In effect, this means that from the gauge-invariant fields $\mb E, \mb B$, the potentials $\mb A, \mb C$ are determined by inverting the curl operator $\nabla \times$. In future work, we plan to make use of the  analogue of the Biot-Savart law for the case of linearized gravity to find a gauge invariant formulation of the Helicity array also in this case.



\subsection*{Acknowledgements} We thank Steffen Aksteiner, Marius Oancea and Kjell Ros\-quist for helpful remarks.

\appendix
\section{Symmetrized energy-momentum tensor and angular momentum in Maxwell theory}\label{app_Symm_emtensor}
The fact that the canonical energy-momentum tensor \eqref{can_em_T} is not symmetric, is intimately related to the (nonconserved) spin tensor as it is obvious from \eqref{EM_nonconservation_eq}. This is the case for all theories with vector (or tensor or spinor)  fields with nonvanishing spin part in angular momentum tensor. It is possible to construct a symmetric energy-momentum tensor, $T_{s}^{\alpha\beta}=T_{s}^{(\alpha\beta)}$, through the Belinfante-Rosenfeld symmetrization procedure (i.e. the addition of a suitable total divergence to the canonical energy–momentum tensor) as
\begin{align}
T_{s}^{\alpha\beta} = T^{\alpha\beta} + \partial_\gamma K^{\alpha\beta\gamma} = \frac12\,(F^{\alpha\gamma}\,F^{\beta}{}_\gamma + G^{\alpha\gamma}\,G^{\beta}{}_\gamma)\ ,\quad \partial_\beta\, T_{s}^{\alpha\beta}=0\,. \label{em-symT}
\end{align}
where $K^{\alpha\beta\gamma}$ is constructed from (nonconserved) spin current tensor\footnote{See the previous section for the expression of and discussion about spin current tensor.} $\tilde S^{\alpha\beta\gamma}$:
\begin{align}
K^{\alpha\beta\gamma}= -\frac12\left(\tilde S^{\alpha\beta\gamma}-\tilde S^{\alpha\gamma\beta}-\tilde S^{\beta\gamma\alpha}\right) = -\frac12\,(A^\alpha F^{\beta\gamma} + C^\alpha G^{\beta\gamma})\,.
\end{align}
The symmetric energy-momentum tensor \eqref{em-symT} is manifestly gauge invariant and contains the momentum density coincident with the energy flux (Poynting vector) $\mb P$. Explicitly, its components are:
\begin{align}
T^{00}_s = \frac12(  E^2 +   B^2) = E\ ,\quad T^{i0}_s = T^{0i}_s = (\mb E \times \mb B)^i = \mb P^i \ ,\quad T^{ij}_s = -\sigma^{ij}
\end{align}
These are the same as components of symmetric energy-momentum tensor derived from the standard (i.e.\! duality \textit{asymmetric}) Lagrangian \eqref{standard_em_L}. 

As a result of this symmetrization procedure, an additional term $P_{\!s}^i=\partial_\gamma K^{i0\gamma}$ contributes to the momentum, modifying it to coincide with the energy flux (i.e the Poynting vector) \cite{Berry2009}:
\begin{align}
\mb P = \mb P_{\!o} + \mb P_{\!s}\ ,\quad \mb P_{\!s} = -\tfrac12\left[\,(\mb E\cdot\nabla)\,\mb A + (\mb B\cdot\nabla)\,\mb C\,\right]\,.
\end{align} 
This is a total derivative in free space and gives a boundary term
\begin{align}
\int_V \mb P_{\!s} \,\ud^3r = -\frac12\,\oint_{\partial V} \left[(\mb A)\,\mb E +(\mb C)\, \mb B\right]. \ud \boldsymbol{\sigma}
\end{align}
in total momentum of the field, thus on whole space for rapidly falling-off fields
\begin{align}
\int \mb P \,\ud^3r = \int \mb P_{\!o} \,\ud^3r \,,
\end{align}
There is a relation between $\mb P_{\!s}$ and spin angular momentum similar to the relation \eqref{orbitalAM} between $\mb P_{\!o}$ and orbital angular momentum, 
\begin{align}
\int_V \mb r \times \mb P_{\!s} \,\ud^3 r = \int_V \mb S \,\ud^3 r   - \oint_{\partial V} \frac12\left[(\mb r\times\mb A) \mb E +(\mb r\times\mb C) \mb B\right]. \ud \boldsymbol{\sigma} \label{rtimesPs}
\end{align}
and in this sense, it can be interpreted as the spin part of the energy flux density.

The angular momentum current constructed from the symmetric energy-momentum tensor $T^{\alpha\beta}_s$ is
\begin{align}
M^{\alpha\beta\gamma}_s = r^\alpha\, T^{\beta\gamma}_s - r^\beta\, T^{\alpha\gamma}_s \ , \quad \partial_\gamma\, M^{\alpha\beta\gamma}_s=0 \label{em_can_an_sym}
\end{align}
which, despite its form, contains both orbital and spin angular momentum and can be separated as
\begin{align}
M_s^{\alpha\beta\gamma} ={}& r^\alpha\, T^{\beta\gamma}_s - r^\beta\, T^{\alpha\gamma}_s = r^\alpha\, T^{\beta\gamma} - r^\beta\, T^{\alpha\gamma} + r^\alpha \partial_\rho K^{\beta\gamma\rho} - r^\beta  \partial_\rho K^{\alpha\gamma\rho} \nonumber\\
& = (r^\alpha T_c^{\beta\gamma} - r^\beta T^{\alpha\gamma}) + (K^{\alpha\gamma\beta} - K^{\beta\gamma\alpha}) + \partial_\rho (r^\alpha K^{\beta\gamma\rho} - r^\beta  K^{\alpha\gamma\rho}) \nonumber\\
={}& \tilde L^{\alpha\beta\gamma} + \tilde S^{\alpha\beta\gamma} + \partial_\rho (r^\alpha K^{\beta\gamma\rho} - r^\beta  K^{\alpha\gamma\rho})
\end{align}
where in the last term, the expression under derivative can be derived as
\begin{align}
r^\alpha K^{\beta\gamma\rho} - r^\beta  K^{\alpha\gamma\rho} = -r^{[\alpha}A^{\beta]}\,F^{\gamma\rho} -r^{[\alpha}C^{\beta]}\,G^{\gamma\rho}\,.
\end{align}
The total angular momentum 3-vector then can be obtained as
\begin{align}
\mb M_s = \mb L + \mb S -\frac12\, \nabla.\,[(\mb r \times \mb A)\mb E + (\mb r \times \mb C)\mb B] = \mb r \times (\mb P_{\!o} + \mb P_{\!s}) = \mb r \times \mb P\,.
\end{align}\\


\newcommand{\mnras}{Monthly Notices of the Royal Astronomical Society }

\newcommand{\arxivref}[1]{\href{http://www.arxiv.org/abs/#1}{{arXiv.org:#1}}}
\newcommand{\prd}{Phys. Rev. D} 
\newcommand{\pra}{Phys. Rev. A} 

\bibliographystyle{abbrv}

\bibliography{references}
\end{document}